\documentclass{article}
\usepackage{graphics}
\begin{document}
\title{Mining for trees in a graph is NP-complete}
\author{Jan Van den Bussche \\ Universiteit Hasselt}
\date{}
\maketitle
\newlength{\natparlength}
The problem of mining patterns in graph-structured data has received
considerable attention in recent years, as it has many interesting
applications in such diverse areas as biology, the life sciences,
the World Wide Web, and social sciences.
Kuramochi en Karypis \cite{kuka_freqpat} identified the following problem as
fundamental to graph mining:
\begin{description}
\item[Problem:] Disjoint Occurrences of a Graph $P$
\item[parameter:] a small graph $P$, called the \emph{pattern}.
\item[input:] a large graph $G$ and a natural number $k$.
\item[decide:] are there $k$ edge-disjoint subgraphs of $G$ that are all
isomorphic to $P$?
\end{description}
Kuramochi and Karypis
relate this problem to Independent Set, a well-known NP-complete
problem \cite{gj_intract}.
They do not go as far, however, as actually proving that Disjoint
Occurrences itself can also be NP-complete.  The purpose of this short
note is to
confirm this, even in the very simple case where $P$ is the tree on four nodes,
consisting of a node with three children.
We denote this tree by $T_3$:
\newcommand{\natparbox}[1]{\settowidth{\natparlength}{#1}%
\parbox{\natparlength}{#1}}
$$ T_3 = \natparbox{\resizebox{3em}{!}{\includegraphics{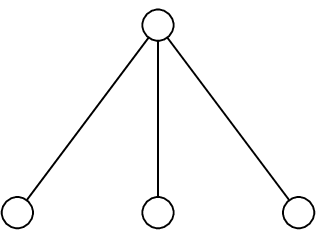}}} $$

As an immediate consequence, the following more general problem, where the
pattern is a tree not fixed in advance but part of the input, is NP-complete
as well:
\begin{description}
\item[Problem:] Disjoint Occurrences of a Tree
\item[input:] a tree $T$, a graph $G$, and a natural number $k$.
\item[decide:] are there $k$ edge-disjoint subgraphs of $G$ that are all
isomorphic to $T$?
\end{description}
Of course, the even more general problem where the pattern is a graph
is already well-known to be NP-complete, as it contains the well-known
NP-complete Subgraph Isomorphism problem
as the special case $k=1$.  We thus see here that restricting to tree patterns
does not lower the worst-case complexity.

The NP-completeness of Disjoint Occurrences of $T_3$ follows immediately from
the NP-completeness of the following problem: (a graph is called
cubic if every node has degree~3, i.e., has edges to precisely 3 other nodes)
\begin{description}
\item[Problem:] Independent Set for Cubic Graphs
\item[input:] a cubic graph $G$ and a natural number $k$
\item[decide:] are there $k$ nodes in $G$ such that no edge runs between these
nodes?
\end{description}
As a matter of fact, for cubic graphs, Indepent Set is
precisely the same problem as Disjoint Occurrences of $T_3$!

To see this, let $G$ be an arbitrary cubic graph.
We call a subgraph of $G$ isomorphic to $T_3$ an \emph{occurrence of $T_3$ in
$G$}. We call the unique node of $T_3$ that has three children the
\emph{center} of $T_3$.  Since $G$ is cubic, we can identify the occurrences
of $T_3$ in $G$ with their centers.  Indeed, every occurrence has a unique
center, and every node is the center of a unique occurrence.
We now easily observe:
\begin{quote}
\emph{Two distinct occurrences of $T_3$, with centers $x$ and $y$, have an
edge in common, if and only if there is an edge between $x$ and $y$.}
\end{quote}
Consequently:
\begin{quote}
\emph{$G$ contains $k$ nodes without edges in between, if and only if
there are $k$ edge-disjoint occurrences of $T_3$ in $G$.}
\end{quote}
In other words, Disjoint Occurrences of $T_3$ is precisely the same problem
as Independent Set, restricted to cubic graphs.

To conclude, we point out that the whole reason for NP-completeness is that we
want to count disjoint occurrences.  Indeed, just counting the occurrences of
a fixed pattern $P$ in a graph can be done in polynomial time.  Unfortunately,
allowing non-disjoint occurrences has problems of its own, as discussed by
Kuramochi and Karypis.

\end{document}